\documentclass{elsart}

\usepackage{graphicx}

\usepackage{amssymb}
\usepackage{amsmath}
\begin{document}

\begin{frontmatter}



\title{Helicity asymmetries in neutrino-nucleus interactions}


\author{P. Lava}\ead{Pascal.Lava@ugent.be},
\author{N. Jachowicz},
\author{M.C. Mart\'{\i}nez},
\author{J. Ryckebusch}

\address{Department of Subatomic and Radiation Physics, Ghent University, Proeftuinstraat 86, B-9000 Gent, Belgium}

\begin{abstract}
We investigate the helicity properties of the ejectile in
quasi-elastic neutrino-induced nucleon-knockout reactions and consider the
$^{12}$C target as a test case. A formalism based on a relativistic
mean-field model is adopted. The influence of final-state interactions
is evaluated within a relativistic multiple-scattering Glauber
approximation (RMSGA) model. Our calculations reveal that the helicity
asymmetries $A_l$ in $A(\overline{\nu},\overline{\nu}'N)$ processes
are extremely sensitive to strange-quark contributions to the weak
vector form-factors. Thereby, nuclear corrections, such as final-state
interactions and off-shell ambiguities in the electroweak current
operators, are observed to be of marginal importance. This facilitates
extracting strange-quark information from the helicity asymmetry
$A_l$.
\end{abstract}

\end{frontmatter}


\section{Introduction}

Parity-violating scattering reactions can be used to probe specific
nucleonic properties which remain concealed in parity-conserving
processes. A subject that has gained wide interest concerns the
contribution of the sea quarks to the nucleon properties such as spin,
charge and magnetic moment.  From the late 1990's on, parity-violating
electron scattering (PVES) has become a tool for hadron physics
research at electron accelerator facilities. Mirror measurements such
as SAMPLE \cite{mueller97} at MIT-Bates, HAPPEX \cite{aniol99,aniol04}
and G0 \cite{armstrong05} at JLAB, A4 \cite{maas04,maas05} at MAMI and
E158 \cite{anthony04} at SLAC aim at probing the strange-quark effects
in proton structure.  In the first place, these collaborations
focus on the strange electric and magnetic
form-factors. Radiative corrections heavily complicate the extraction
of the strange axial form-factor $g_A^s$ from the data. In the
analysis of the parity-violating asymmetry observed in the PVES
experiments, one estimates the effect of $g_A^s$ relying on results of
deep-inelastic double-polarized scattering experiments
\cite{ashman89,adeva98,anthony00,airapetian04}. The abovementioned
PVES programs triggered many theoretical studies of the strangeness
magnetic moment and charge radius.  These calculations are performed
in a rich variety of hadron models, yielding predictions for the
strangeness parameters covering a wide range of values
\cite{jaffe89,musolf94,koepf92,cohen93,musolf97,kim95,park91,musolf97prd,silva01}.
A recent review of the theoretical and experimental status can be
found in Ref. \cite{musolf05}.

An alternative method of addressing the strangeness content of the
nucleon is by means of neutrino-nucleus scattering. In contrast to
PVES experiments, in extracting $g_A^s$ no radiative corrections need
to be applied. Data for $(\nu,\nu'N)$ and
$(\overline{\nu},\overline{\nu}'N)$ elastic scattering cross sections
were collected at BNL \cite{ahrens87}.  As carbon was used as target
material, an accurate understanding of nuclear corrections is a
prerequisite for reliably extracting the strange-quark matrix elements
from the data.  Examples of relativistic studies which address the
issue of computing the nuclear corrections, are the relativistic Fermi
gas (RFG) model of Ref.~\cite{horo93} and the relativistic
distorted-wave impulse approximation (RDWIA) models of
Refs.~\cite{alberico02,meucci04}.  As absolute cross-section
measurements involving neutrinos are challenging, a lot of effort has
been devoted to the study of cross-section ratios. Examples include
the ratio of proton-to-neutron knockout in neutral-current (NC)
neutrino-nucleus interactions
\cite{alberico02,meucci04,garvey93,vanderventel04}, the ratio of NC to
charged-current (CC) cross sections \cite{pate05,vanderventel05} and
the ratio of NC to CC neutrino-antineutrino asymmetries
\cite{alberico02}.  For these ratios, the effects of nuclear
corrections nearly cancel, facilitating the extraction of possible
strange-quark contributions.  Other observables which do not require
absolute cross-section measurements are polarization
asymmetries. Recently, the helicity asymmetry $A_l$ was put forward as
a potential tool to discriminate between neutrinos and antineutrinos
in NC neutrino-induced nucleon-knockout reactions off nuclei
\cite{jachowicz04,jachowicz05}.  In this paper, we wish to point out
that the quantity $A_l$ is also very sensitive to sea-quark
contributions to the vector form-factors. This is an interesting
insight, since the ratios discussed in
Refs. \cite{alberico02,meucci04,garvey93,vanderventel04,pate05,vanderventel05}
focus on effects stemming from $g_A^s$. Indeed, in those ratios the
axial part largely overshoots contributions of the vector
form-factors.  Often, the extraction of physical information from
observables involving nuclei suffer from an incomplete knowledge of
medium effects.  In this paper it is shown that $A_l$ remains
relatively free of these ambiguities.  We adopt a relativistic
framework to compute the nuclear medium effects on $A_l$.

The outline of this paper is as follows. In Sec. \ref{sec:two} we
present the relativistic multiple-scattering Glauber approximation
(RMSGA) formalism for the description of the helicity asymmetry within
NC neutrino-nucleus scattering processes. In Sec. \ref{sec:three} we
present our results for $A_l$, and pay particular attention to the
influence of medium corrections and strangeness contributions. In
Sec. \ref{sec:four} we summarize our findings.

\section{Formalism}

\label{sec:two}

The expressions for neutrino and antineutrino quasi-elastic
neutral-current (NC) reactions from nuclei which result in one emitted
nucleon,
$\nu(\overline{\nu})+A\rightarrow\nu(\overline{\nu})+N+(A-1)$, are
derived in Ref. \cite{martinez05}. Within the one-boson exchange
approximation the one-fold differential cross section reads
\begin{eqnarray} 
\label{eq:intcross}
\frac{d\sigma}{dT_{N}}  & = & \frac{M_NM_{A-1}}{(2\pi)^3M_A}4\pi^2 
 \int\sin{\theta_l}d\theta_l\int \sin{\theta_N}d\theta_N \; \nonumber\\ & & \times
 k_N f_{rec}^{-1} \sigma_M[v_LR_L + v_TR_T + hv_{T'}R_{T'}] \;. 
\end{eqnarray} 
In this expression, $M_N$ ($T_N$) represents the mass (kinetic energy)
of the ejectile, while $M_A$ ($M_{A-1}$) refers to the mass of the
target (residual) nucleus. The outgoing nucleon momentum is
$\vec{k}_N$, $f_{rec}$ is the recoil factor and $\sigma _M$ a
Mott-like cross section. The direction of the scattered lepton
(outgoing nucleon) is fixed by the angles $\Omega_l$ ($\Omega_N$). In
Eq.~(\ref{eq:intcross}), the helicity is $h \mbox{=} -1$ ($h \mbox{=}
+1$) for neutrinos (antineutrinos). Expressions for the kinematic
factors $v_L,v_T,v_{T'}$ and the structure functions $R_L,R_T,R_{T'}$
can be found in Ref. \cite{martinez05}. The latter ones embody the
effects of the nuclear dynamics. In the calculation of the responses, the
basic quantity to be computed is the transition matrix element
$\langle J^{\mu} \rangle$.  Adopting the impulse approximation and an
independent-nucleon picture, $\langle J^{\mu} \rangle$ can be
expressed as
\begin{equation}
\label{eq:relcurrent}
  \langle J^{\mu}
  \rangle = \int
  d\vec{r} \; \overline{\phi}_F(\vec{r})\widehat{J}^{\mu}(\vec{r})e^{i\vec{q}.\vec{
      r}}\phi_{B}(\vec{r}) \; ,
 \end{equation}
where $\phi_{B}$ and $\phi_F$ are relativistic bound-state and
scattering wave-functions, and $\widehat{J}^{\mu}$ is the electroweak
current operator. The wave functions $\phi_{B}$ are
obtained within the Hartree approximation to the $\sigma$-$\omega$
model \cite{serot86}.  

For a
free nucleon, the one-body vertex function $J^{\mu}$ can be expressed
in several equivalent forms of which some of the more frequently
used ones read \cite{Forest1983}
\begin{subequations}
\label{eq:current}
\begin{eqnarray}
J^{\mu}_{cc1} &=& G_M^Z(Q^2)\gamma^{\mu} -
\frac{\kappa}{2M_N}F_2^Z(Q^2)(K_i^{\mu} + K_f^{\mu}) +
G_A(Q^2)\gamma^{\mu}\gamma_5 \; ,
\\
\label{eq:cc2}
J^{\mu}_{cc2} &=& F_1^Z(Q^2)\gamma^{\mu} +
i\frac{\kappa}{2M_N}F_2^Z(Q^2)\sigma^{\mu\nu}q_{\nu} +
G_A(Q^2)\gamma^{\mu}\gamma_5 \; ,
\\
J^{\mu}_{cc3} &=& \frac{1}{2M_N}F_1^Z(Q^2)(K_i^{\mu} + K_f^{\mu}) +
i\frac{1}{2M_N}G_M^Z (Q^2) \sigma^{\mu\nu}q_{\nu} +
G_A(Q^2)\gamma^{\mu}\gamma_5 \;,
\end{eqnarray} 
\end{subequations}
with $q^{\mu} = (\omega, \vec{q})$ and $Q^2 = -q_{\mu}q^{\mu}$ the
four-momentum transfer.
The relation between the weak Sachs electric and magnetic form-factors
$G_E^Z$ and $G_M^Z$ and the weak Dirac and Pauli form-factors $F_1^Z$
and $F_2^Z$, is established in the standard fashion. However, when considering
off-shell nucleons embedded in a nuclear medium, the above
vertex functions can no longer be guaranteed to produce identical
results.  This elusive feature is known as the Gordon ambiguity and is
a source of uncertainties when performing calculations involving
finite nuclei \cite{Forest1983,Pollock96,Kelly97}.

The weak vector form-factors $F_1^Z$ and $F_2^Z$ can
be expressed in terms of the electromagnetic form-factors for
protons $(F_{i,p}^{EM})$ and neutrons $(F_{i,n}^{EM})$ by the
conserved vector current (CVC) hypothesis
\begin{eqnarray}
\label{eq:vectorform}
F_{i}^Z &=& \left(\frac{1}{2}-\sin^2{\theta_W}\right)\left(F_{i,p}^{EM}
- F_{i,n}^{EM}\right)\tau_3 \nonumber \\ &-& \sin^2{\theta_W}\left(F_{i,p}^{EM} +
F_{i,n}^{EM}\right) - \frac{1}{2} F_i^s \; \; \; \; (i=1,2) \; ,
\end{eqnarray}
with $\sin^2{\theta_W} = 0.2224$ the Weinberg angle and $F_i^s$
quantifying the effect of the strange quarks. The isospin
operator $\tau_3$ equals $+1$ ($-1$) for protons (neutrons).  For
long, the accumulated data pointed towards electromagnetic
form-factors of the nucleon whose $Q^2$- dependence can be well
described in terms of a dipole parameterization. Traditionally, these
data were obtained by means of a Rosenbluth separation of elastic
$p(e,e')p$ scattering measurements. New data based on polarization
transfer measurements $p(\vec{e},e')\vec{p}$ \cite{jones00,gayou02}
revealed a quite different picture for $Q^2 \ge 1$ (GeV/c)$^2$.  The
discrepancy between the electromagnetic form-factors obtained with the
two techniques is an unresolved issue, but two-photon exchange
processes have been shown to play a major role
\cite{guichon03,blunden03}.

The axial form-factor
can be parameterized as
\begin{equation}
G_A (Q^2) =  - \frac{(\tau_3g_A - g_A^s)}{2} G(Q^2), 
\label{eq:ga}
\end{equation}
with $g_A \mbox{=} 1.262$,  $G=(1+Q^2/M^2)^{-2}$ with $M=1.032$
GeV, and $g_A^s$ the axial strange-quark contribution.

The remaining ingredient entering Eq.~(\ref{eq:relcurrent}) is the
relativistic scattering wave-function $\phi_F$ for the emitted nucleon. 
We incorporate FSI effects in a relativistic version of the Glauber
model which has been dubbed RMSGA \cite{ryckebusch03}. The RMSGA
represents a multiple-scattering extension of the eikonal
approximation and the effects of FSI are directly computed from the
elementary nucleon-nucleon scattering data through the introduction of
a profile function. The Glauber method postulates linear trajectories
for the ejectile and frozen spectator nucleons in the residual
nucleus, resulting in a scattering wave-function of the form
\begin{equation}
\label{eq.:transition}
\phi_F({\vec{r}}) \equiv  \mathcal{G}
\left(\vec{b} (x,y) ,z \right)\phi _{{k}_N, \; s_N}(\vec{r}) \; ,
\end{equation}
where $\phi_{{k}_N, \; s_N}$ is a relativistic plane-wave.  The impact
of FSI mechanisms on the scattering wave function is contained in the
scalar Dirac-Glauber phase $\mathcal{G}(\vec{b},z)$
\cite{ryckebusch03}. The limit of vanishing FSI, i.e. the relativistic
plane-wave impulse approximation (RPWIA), is reached by putting this
phase to unity.  The RMSGA model was successfully tested against
exclusive $A(e,e'p)$ data \cite{lava03,lava04}. The validity of the
RMSGA model in the low-energy regime was tested by comparing its
predictions to results from an RDWIA calculation
\cite{martinez05}. Satisfying RMSGA results down to nucleon kinetic
energies of $250$ MeV were found.

The longitudinal polarization asymmetry $A_l$, which will be the
object of discussion in this paper,  is defined as the
difference in yield for the two possible helicity states of the
ejectile, normalized to the total differential nucleon knockout cross
section:
\begin{equation}
\label{eq:polasym}
A_{l}(T_N) = \frac{\frac{d\sigma}{dT_{N}}(h_N = +1) - 
\frac{d\sigma}{dT_{N}}( h_N = -1)} 
{\frac{d\sigma}{dT_{N}}(h_N = +1) +
\frac{d\sigma}{dT_{N}}(h_N = -1)} \; ,
\end{equation}
where $h_N = \frac{\vec{\sigma}_N \cdot \vec{k}_N } {\mid \vec{k} _N
\mid } $ denotes the helicity of the ejected nucleon. 

\section{Results}

\label{sec:three}
The primary goal of this paper is to scrutinize the impact of
strange-quark contributions on the observable $A_l$. First, we wish to
determine the degree to which $A_l$ is affected by variations in the
parameterizations for the electromagnetic form-factors and typical
medium effects like FSI and off-shell ambiguities. We consider the
$^{12}$C target as a test case. We take RPWIA calculations as baseline
results, with dipole form-factors and the current operator in its
$CC2$ form of Eq.~(\ref{eq:cc2}). 

\begin{figure}[p]
  \begin{center}
    \caption{ The helicity asymmetry as a function of $T_p$ for
  proton knockout from $^{12}$C at six beam energies. The left (right)
  panel is for neutrinos (antineutrinos).}
    \label{fig:ener}
    \includegraphics[width=0.8\textwidth]{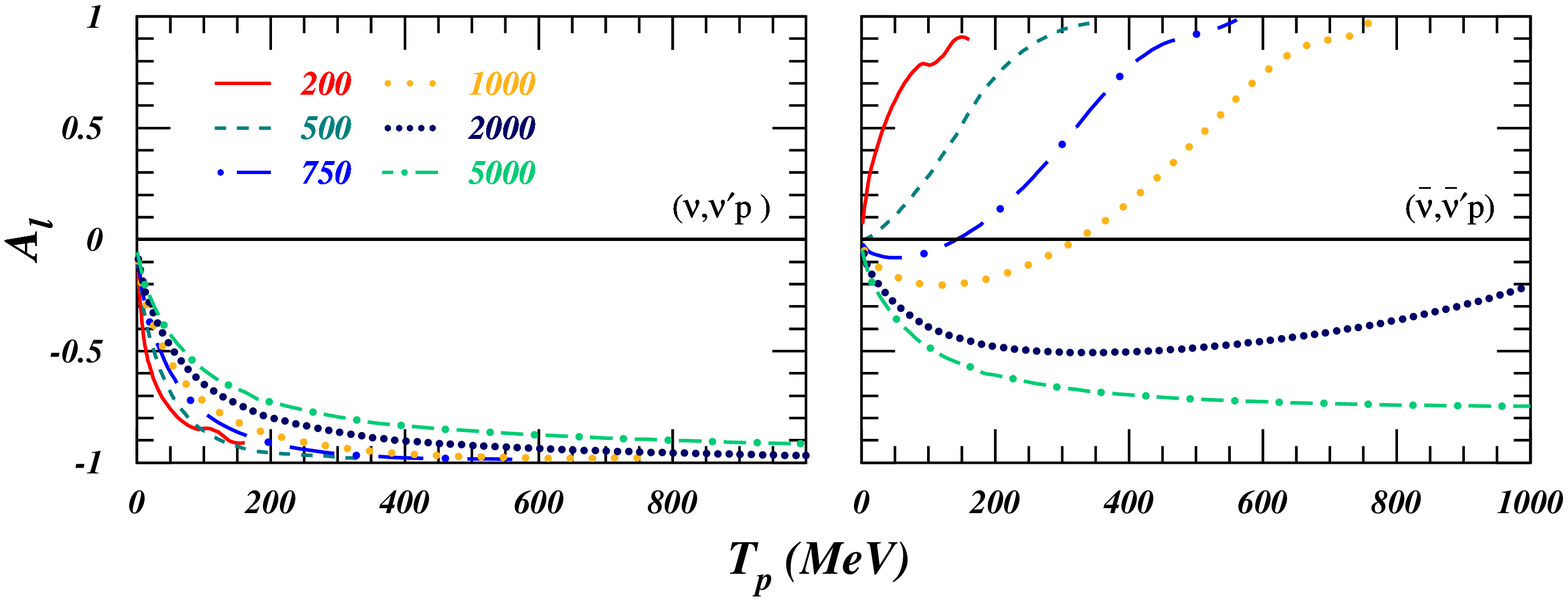}
  \end{center}
\end{figure}

As mentioned earlier, in Ref.~\cite{jachowicz04} the helicity
asymmetry $A_l$ was put forward as a lever to discriminate
between neutrinos and antineutrinos in NC reactions on nuclei.
Predictions for this asymmetry were obtained in a
non-relativistic plane-wave impulse approximation and results up to 
beam energies of $500$ MeV were presented. In the GeV energy domain,
any realistic model requires the inclusion of relativistic effects. 
In Fig.~\ref{fig:ener}, we show the RPWIA predictions for $A_l$ for
beam energies ranging from $200$ to $5000$ MeV. Clearly, up to lepton
energies of 1 GeV, the $A_l$ has an opposite sign for $A(\nu,\nu 'N)$
and $A( \overline{\nu}, \overline{\nu} 'N)$.  Apparently, the
discriminative power of $A_l$ dwindles when higher beam energies are
considered. The antineutrino asymmetry evolves from a dominance of
$h_N = +1$ contributions at ``low'' beam energies to a supremacy of
$h_N = -1$ ones at ``higher'' energies. This can be
attributed to the role played by the $G_AF_2^Z$ interference
contribution.

\begin{figure}[p]
  \begin{center}
    \caption{The effect of FSI mechanisms on the helicity asymmetry at
    $500$ MeV and $1000$ MeV beam energies. The solid (dashed) line
    shows the RPWIA (RMSGA) predictions.}
    \label{fig:fsi}
    \includegraphics[width=0.8\textwidth]{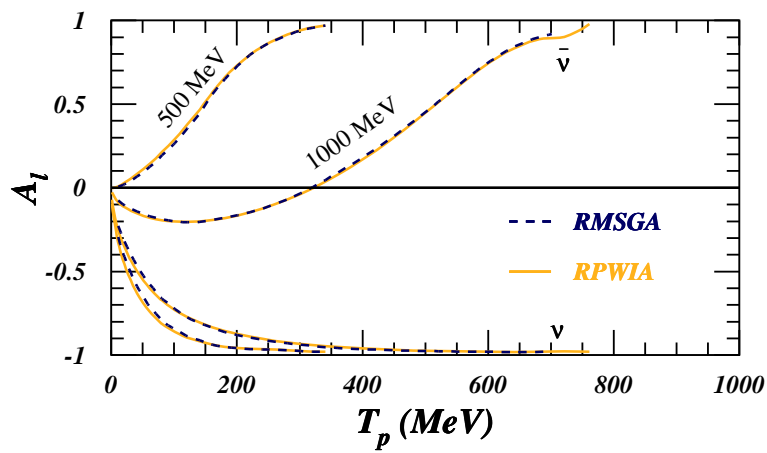}
  \end{center}
\end{figure}

None of the results for $A_l$ shown so far, including those of
Refs.~\cite{jachowicz04,jachowicz05}, did account for the effects of
FSI. We compute these within the RMSGA
model. Fig.~\ref{fig:fsi} displays the effect at beam energies of
$500$ MeV and $1000$ MeV. As can be appreciated, the global influence of
FSI mechanisms on $A_l$ is almost negligible. In the ratio $A_l$ a
strong cancellation of FSI is noticed, even at relatively low
ejectile kinetic energies. This should not be considered as a trivial
result, since FSI do play an important role in the corresponding
inclusive cross sections \cite{martinez05}. Henceforth, we will
concentrate on an incoming energy of $\varepsilon \mbox{=} 1000$ MeV.
At this energy, the neutrino scattering process can be expected to be
dominated by the quasi-elastic contribution.

\begin{figure}[p]
  \begin{center}
    \caption{The helicity asymmetry $A_l$ as a function of the proton
    kinetic energy at $\varepsilon \mbox{=} 1000$ MeV as computed in an
    RPWIA approach. The left panel illustrates the effects stemming
    from the ambiguities in the electromagnetic form-factors: the
    solid (dashed) line shows the RPWIA results obtained with the
    dipole (BBA-2003) parameterization. In the right panel the role of
    the off-shell ambiguities is studied. The solid, dashed, dot-dashed
    curves are obtained with the CC2, CC1 and CC3 prescription, respectively. }
    \label{fig:ff}
    \includegraphics[width=0.8\textwidth]{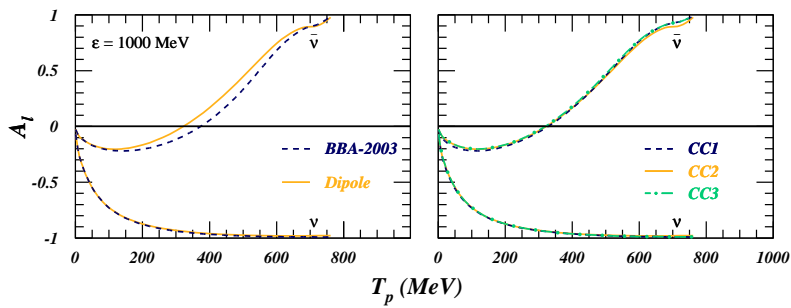}
  \end{center}
\end{figure}

Another possible source of uncertainty when determining $A_l$ may be
the insufficient knowledge regarding the electromagnetic form-factors
of the proton. We performed calculations with two parameterizations:
the standard dipole form and the BBA-2003 parameterization of
Ref.~\cite{budd03}. As becomes clear from the left panel of
Fig.~\ref{fig:ff}, both produce comparable results.
Therefore, all forthcoming results use the traditional dipole
for $G_E^{EM}$ and $G_M^{EM}$. We also wish to
estimate the role of off-shell ambiguities on the computed $A_l$
values.  To that purpose we performed calculations with all current
operators of Eq.~(\ref{eq:current}). As fig.~\ref{fig:ff} shows that
all these current operators produce more or less equivalent results,
the sensitivity of $A_l$ to off-shell ambiguities is minor.

The helicity asymmetry $A_l$ emerges as a robust observable, which is
not burdened by a large sensitivity to medium corrections.  So far, we
neglected strangeness contributions to the weak vector and axial
form-factors of Eqs.~(\ref{eq:vectorform}) and (\ref{eq:ga})
($F_1^s=F_2^s=g_A^s=0$). To quantify the impact of the axial
strangeness contribution on $G_A$, we adopt the value $g_A^s =
-0.19$. This value was extracted from an $SU(3)$-based analysis of
deep inelastic double-polarized scattering experiments
\cite{ashman89}.  In addition to sea-quark effects in the axial
current, there can be contributions to the Dirac and Pauli vector
form-factors.  A three-pole ansatz of Forkel et al. \cite{forkel97}
resulted in the following parameterization
\begin{equation}
F_1^s = \frac{1}{6} \frac{-r_s^2Q^2}{(1 + Q^2/M_1^2)^2} \; ,
\end{equation}
\begin{equation}
F_2^s = \frac{\mu_s}{(1 + Q^2/M_2^2)^2} \; ,
\end{equation}
with $M_1 \mbox{=} 1.3$ GeV and $M_2 \mbox{=} 1.26$ GeV \cite{forkel97}. The
$r_s^2$ and $\mu _s$ predicted by various hadronic
structure models are summarized in Table \ref{table1}.  
The list is not exhaustive. There is a tendency towards 
a mildly negative strangeness magnetic moment ($\mu_s \approx -0.3
$~$\mu_N$), and a small negative strangeness radius ($r_s^2 \approx -0.01
$~$fm^2$). However, all PVES experiments
performed so far suggest a positive value for $\mu_s$. In our
investigations we will use the $r_s^2$ and $\mu _s$ from the vector meson
dominance (VMD), the
K$\Lambda$, the Nambu-Jona-Lasinio (NJL)  and the chiral quark soliton
(CQS(K)) model.  These values are selected as
we find them representative for the full range of values
regarding the strangeness parameters.

\begin{figure}[p]
  \begin{center}
    \caption{ Sensitivity of the proton Pauli and Dirac neutral-current
    vector form-factors to strange-quark contributions. The solid line
    represents the form factors in the absence of any strangeness
    contribution. The dashed, dot-dashed, long-dotted and short-dotted
    curves include non-zero strangeness contributions according to the
    prediction of four different hadron models: VMD \cite{jaffe89},
    K$\Lambda$ \cite{musolf94}, NJL \cite{kim95} and CQS model
    \cite{silva01} respectively.}
    \label{fig:Fp}
    \includegraphics[width=0.8\textwidth]{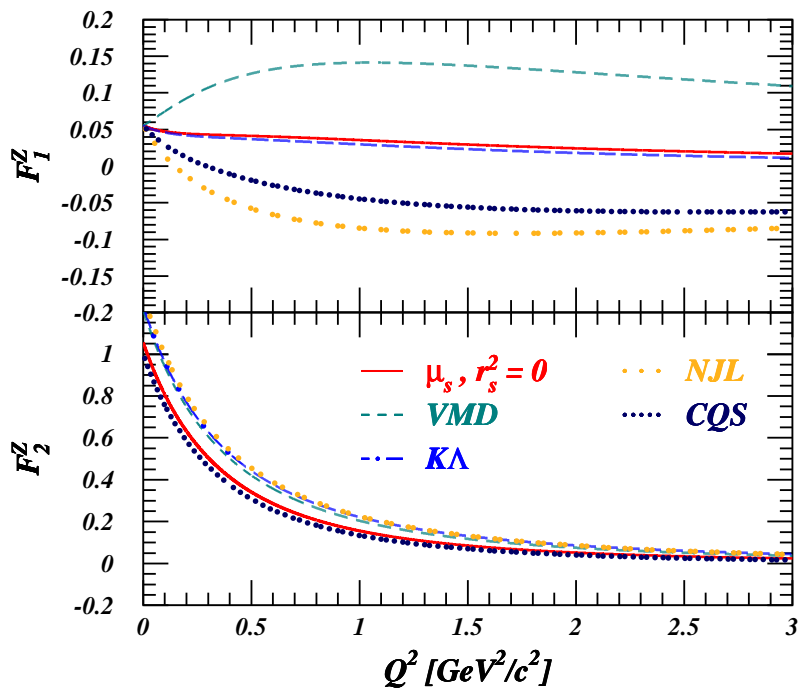}
  \end{center}
\end{figure}

In Fig. \ref{fig:Fp}, the proton Dirac $F_1^Z$ and Pauli $F_2^Z$ NC
form-factors are shown for various strangeness parameters $F_1^s$ and
$F_2^s$. This figure reveals that mainly $F_1^Z$ is affected. Indeed,
the VMD model predicts a Dirac form-factor that overshoots the
non-strangeness and other model predictions by a factor of about
three. In addition, a sign switch in $F_1^Z$ appears in the ``NJL''
and ``CQS'' models. Variations in the Pauli form-factor however are
less pronounced due to its large absolute value. Thus, one can expect
that mainly variations in $r_s^2$ will be reflected in the helicity
asymmetry.

\begin{figure}[p]
  \begin{center} 	
   \caption{Influence of sea-quarks on the helicity asymmetry at
   $\varepsilon \mbox{=} 1000$ MeV. The left panel shows the asymmetry
   for antineutrino-induced proton knockout on $^{12}$C, whilst the
   right one shows the asymmetry for antineutrino-induced
   neutron knockout. The solid curve represents the RPWIA results
   without strangeness. The other curves correspond to different
   strangeness parameterizations: $g_A^s \mbox{=} -0.19$ (dashed), VMD
   (long dot-dashed) \cite{jaffe89}, K$\Lambda$ (long-dotted)\cite{musolf94}, NJL (short-dotted)\cite{kim95} and CQS
   (short dot-dashed)\cite{silva01}.}
\label{fig:asymstrange}
    \includegraphics[width=0.8\textwidth]{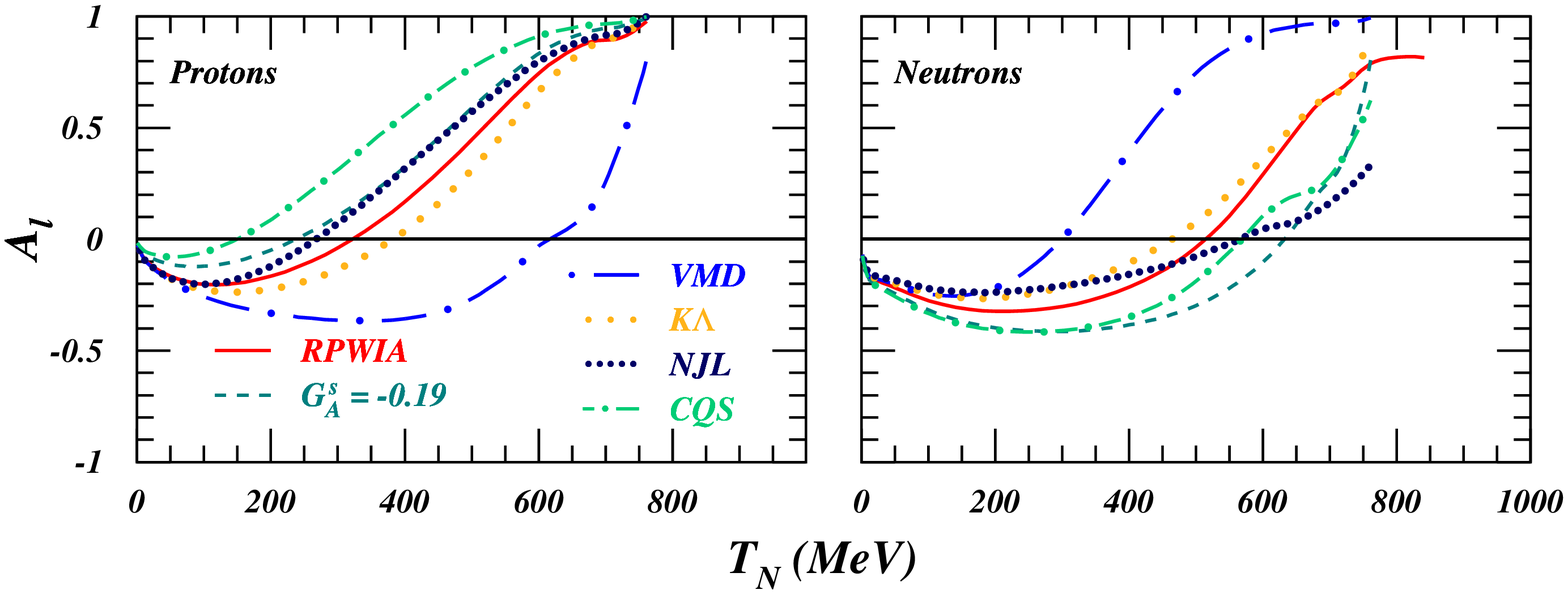}
  \end{center} 
\end{figure}

Fig. \ref{fig:asymstrange} shows our predictions for the helicity
asymmetry at $\varepsilon \mbox{=} 1000$ MeV for both proton and
neutron knockout. The results contained in Figs. \ref{fig:ener},
\ref{fig:fsi} and \ref{fig:ff} reveal that neutrinos are extremely
selective with respect to the helicity of the ejectile. As a
consequence, one can expect that any strangeness contribution will
nearly cancel in the ratio of Eq.~(\ref{eq:polasym}). The helicity
preference is less outspoken for antineutrinos. Hence, antineutrinos
represent a better lever than neutrinos when it comes to probing
strange-quark contributions through the observable $A_l$. For both
protons and neutrons, the introduction of a non-zero $g_A^s$ does not
substantially alter the baseline results (denoted as RPWIA in the
figure).  The introduction of non-zero strangeness radius and magnetic
moment, on the other hand, seriously affects the ratio between $h_N =
+1$ and $h_N= -1$ ejectiles. The largest deviations emerge using the
predictions of the VMD model ($r_s^2 > 0$).  In any case, the overall
impact of $F_1^s$ and $F_2^s$ on the helicity asymmetry is
substantially larger than the effect caused by FSI mechanisms,
off-shell ambiguities and $g_A^s$. As can be inferred from
Fig.~\ref{fig:asymstrange}, the strange contribution to the weak
vector form-factors has a comparable impact on the $A_l$ for protons
and neutrons, but acts in opposite directions. This is another
illustration of the well-known feature that in hunting sea-quarks in
neutrino or PVES reactions, it is essential to discriminate between
protons and neutrons. Indeed, the effects stemming from the sea quarks
tend to cancel when summing over the proton and neutron
cross-sections.

\begin{figure}[p]
  \begin{center} 	
   \caption{The helicity asymmetry for antineutrino-induced
   proton knockout at $\varepsilon \mbox{=} 1000$ MeV. The solid line
   shows the RPWIA predictions with $g_A^s \mbox{=} -0.19 $. The left
   (right) panel gives the influence of varying strangeness radius (magnetic moment).  }
\label{fig:asymrs}
    \includegraphics[width=0.8\textwidth]{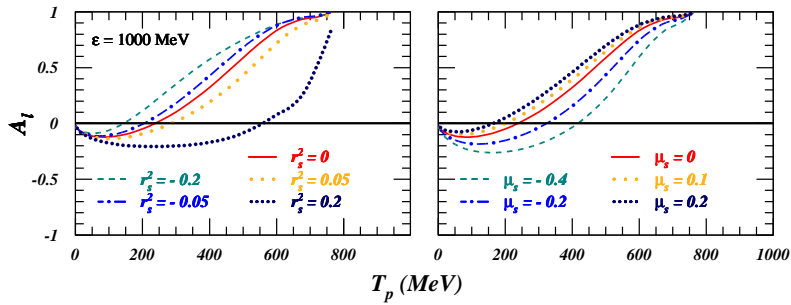}
  \end{center} 
\end{figure}

The effect of  varying $r_s^2$ and $\mu_s$ separately is studied in
Fig. \ref{fig:asymrs}. In the right panel, we investigate the effect
of varying $\mu_s$ at $r_s \mbox{=} 0$. Not surprisingly, we see that
changing $\mu _s$ has a relatively modest effect on the helicity
asymmetry. Indeed, from Fig. \ref{fig:Fp} one can infer small
variations in the  weak Pauli form-factor. The largest changes in $A_l$
are induced by variations in the strangeness radius $r_s^2$.

\section{Conclusions}

\label{sec:four}
We have studied the helicity properties of the ejectile in
quasi-elastic neutrino-induced nucleon-knockout reactions. Results for $^{12}$C
have been presented for a wide range of (anti)neutrino energies. In
$A(\overline{\nu},\overline{\nu} 'N)$ processes, the helicity
asymmetry is found to be very sensitive to strangeness contributions
in the weak Dirac and Pauli form-factors. Extraction of strange-quark
information from $A_l$ is facilitated by the fact that nuclear
structure effects, such as final-state interactions, form-factor
parameterization and off-shell ambiguities do not affect the
asymmetry. Moreover, strangeness variations in the vector form-factors largely
overshoot effects stemming from the axial part.

\section*{Acknowledgements}
The authors are grateful to the Fund for Scientific Research (FWO)
Flanders, to the University Research Board (BOF) and to the Ministerio
de Educacion y Ciencia (MEC) of Spain (project BFM 2003-04147-C02-01)
for finantial support. M.C.M. would like to acknowledge a postdoctoral
fellowship from the MEC.



\begin{table}[p]
\caption{Predictions for $r_s^2$ and $\mu_s$ in various hadron models.}
\begin{tabular}{llll}
\hline
Model  & Ref.  &  $\mu_s(\mu_N)$ & $r_s^2$(fm$^{2}$)\\
\hline
$VMD$ &  \cite{jaffe89} &  -0.31 &  0.16 \\
$K\Lambda$ &  \cite{musolf94} &  -0.35 & -0.007 \\
$CBM$  & \cite{koepf92}  &-0.1  & -0.011 \\
Hybrid  & \cite{cohen93}  & -0.3 & -0.025 \\
Chiral Quark  & \cite{musolf97}  & -0.09  &-0.035 \\
NJL  & \cite{kim95}  & -0.45 &  -0.17 \\
Skyrme  & \cite{park91}  & -0.13 -- -0.57  & -0.1 -- -0.15 \\
Disp. Rel.  & \cite{musolf97prd}  &-0.28  & 0.42 \\
CQS ($\pi$) & \cite{silva01}  & 0.074  & -0.22 \\
CQS (K)  & \cite{silva01}  & 0.115  & -0.095 \\ 
\hline
\end{tabular}
\label{table1}
\end{table}

\end{document}